\documentclass[sigconf]{acmart}

\AtBeginDocument{%
  \providecommand\BibTeX{{%
    \normalfont B\kern-0.5em{\scshape i\kern-0.25em b}\kern-0.8em\TeX}}}
\usepackage{amsmath}
\usepackage{amssymb}
\usepackage{amsfonts}
\usepackage{algpseudocode}
\usepackage{algorithm}
\usepackage{algorithmicx}

\usepackage{graphicx}
\usepackage{textcomp}
\usepackage{url}

\usepackage{microtype}
\usepackage{multirow}
\usepackage[inline, shortlabels]{enumitem} 

\usepackage{booktabs}
\usepackage{caption,subcaption}

\begin{document}

\title{Familiarity-based Collaborative Team Recognition in Academic Social Networks}


\author{Shuo Yu}
\affiliation{%
\department{School of Software}
\institution{Dalian University of Technology}
\city{Dalian}
\postcode{116620}
\country{China}}
\email{y\_shuo@outlook.com}

\author{Feng Xia}
\affiliation{%
\department{School of Engineering, IT and Physical Sciences}
\institution{Federation University Australia}
\city{Ballarat}
\postcode{VIC 3353}
\country{Australia}}
\email{f.xia@ieee.org}

\author{Chen Zhang}
\affiliation{%
\department{School of Software}
\institution{Dalian University of Technology}
\city{Dalian}
\postcode{116620}
\country{China}}
\email{chen.zhang07@outlook.com}

\author{Haoran Wei}
\affiliation{%
\department{School of Software}
\institution{Dalian University of Technology}
\city{Dalian}
\postcode{116620}
\country{China}}
\email{willieying@outlook.com}

\author{Kathleen Keogh}
\affiliation{%
\department{School of Engineering, IT and Physical Sciences}
\institution{Federation University Australia}
\city{Ballarat}
\postcode{VIC 3353}
\country{Australia}}
\email{k.keogh@federation.edu.au}

\author{Honglong Chen}
\affiliation{%
\department{College of Control and Science Engineering}
\institution{China University of Petroleum}
\city{Qingdao}
\postcode{266580}
\country{China}}
\email{chenhl@upc.edu.cn}

\renewcommand{\shortauthors}{Yu, et al.}

\begin{abstract}
 Collaborative teamwork is key to major scientific discoveries. However, the prevalence of collaboration among researchers makes team recognition increasingly challenging. Previous studies have demonstrated that people are more likely to collaborate with individuals they are familiar with. In this work, we employ the definition of familiarity and then propose MOTO (faMiliarity-based cOllaborative Team recOgnition algorithm) to recognize collaborative teams. MOTO calculates the shortest distance matrix within the global collaboration network and the local density of each node. Central team members are initially recognized based on local density. Then MOTO recognizes the remaining team members by using the familiarity metric and shortest distance matrix. Extensive experiments have been conducted upon a large-scale data set. The experimental results show that compared with baseline methods, MOTO can recognize the largest number of teams. The teams recognized by MOTO possess more cohesive team structures and lower team communication costs compared with other methods. MOTO utilizes familiarity in team recognition to identify cohesive academic teams. The recognized teams are in line with real-world collaborative teamwork patterns. Based on team recognition using MOTO, the research team structure and performance are further analyzed for given time periods. The number of teams that consist of members from different institutions increases gradually. Such teams are found to perform better in comparison with those whose members are from the same institution.
\end{abstract}





\maketitle

\section{Introduction}
\label{sec:1}
Scientific discovery in the 21st century relies on global interactions and collaborations between colleagues. Individual research has increasingly been replaced by collaborative teamwork. Scientific collaboration has been regarded as one of the most effective ways to solve complicated scientific research problems~\cite{vsubelj2019convexity,wang2020early}. Information technology has greatly facilitated communication between scholars. A variety of tools and methods enable collaboration between scholars so that the team can be distributed across various locations. Understanding and being aware of the inner patterns of collaborative academic teamwork may improve team efficiency and the quality of scientific research. It is also possible to gain insights by studying the inner patterns of collaborative teams at both micro and meso-levels. The scale of collaborative teamwork as well as the quality of teamwork collaboration have increased gradually~\cite{hall2018science}. A branch of science entitled “Science of Scientific Team Science (SSTS)”~\cite{yu2019science} has been proposed to study collaborative teamwork, with the aims of enhancing scientific collaboration within teams and improving transdisciplinary research.

Studies of teamwork in management science and psychological science exist, and most of these focus on collaborative teamwork patterns~\cite{ribeiro2018growth,van2018developing}, team recognition~\cite{yu2017team} and team performance enhancement~\cite{kim2017makes}. Other studies include those that investigate attitudes to teamwork. Network science approaches have been proposed to effectively analyze, study, and explore collaborative teamwork patterns~\cite{silva2016using}. Many large-scale academic networks based on large collaboration and citation relationships can be constructed from easily accessed digital libraries such as DBLP, CiteSeerX and MAG. Academic networks provide more complex opportunities for scientific team research~\cite{zhang2018ranking} than traditional research methods that commonly employ case studies or small samples. However, collaborative teamwork data are contained in these networked data in implicit ways, which makes it difficult for scholars to directly employ collaborative team data in their studies~\cite{kong2019academic}. Therefore, one of the most fundamental research problems is to automate the recognition of collaborative teams within academic networks.

Related studies focusing on team recognition have also been conducted. Yu et al.~\cite{yu2017team} propose a network-based approach to identify collaborative teams from academic networks. They propose an index entitled CII (Collaboration Intensity Index) that qualifies collaboration intensity. CII is employed to identify if a certain relationship belongs to a team. Some related studies also use community detection methods to identify collaborative teams~\cite{javed2018community,chakraborty2015formation}. Some community detection methods will recognize clusters of scholars. However, communities are generally identified on a broader basis than collaborative teams. Though some community detection methods can recognize fine-grained communities, scale is not the only difference between academic communities and collaborative teams. Stability is also a feature of significance in the academic collaborative team. It has been shown that a team will achieve better performance when the collaboration patterns among members tend to be stable~\cite{hung2017exploring,yu2019team}. To be specific, members would like to work with those they are more familiar with. A relatively stable communication pattern as well as work pattern will help improve teamwork performance~\cite{kulp2019comparing}. It is therefore important to consider familiarity as a contribution to stability when recognizing collaborative teams. It is difficult, however to qualify the degree of familiarity between members, especially in large scale academic networks.

\begin{figure}[tb]
  \centering
  \includegraphics[width=0.95\linewidth]{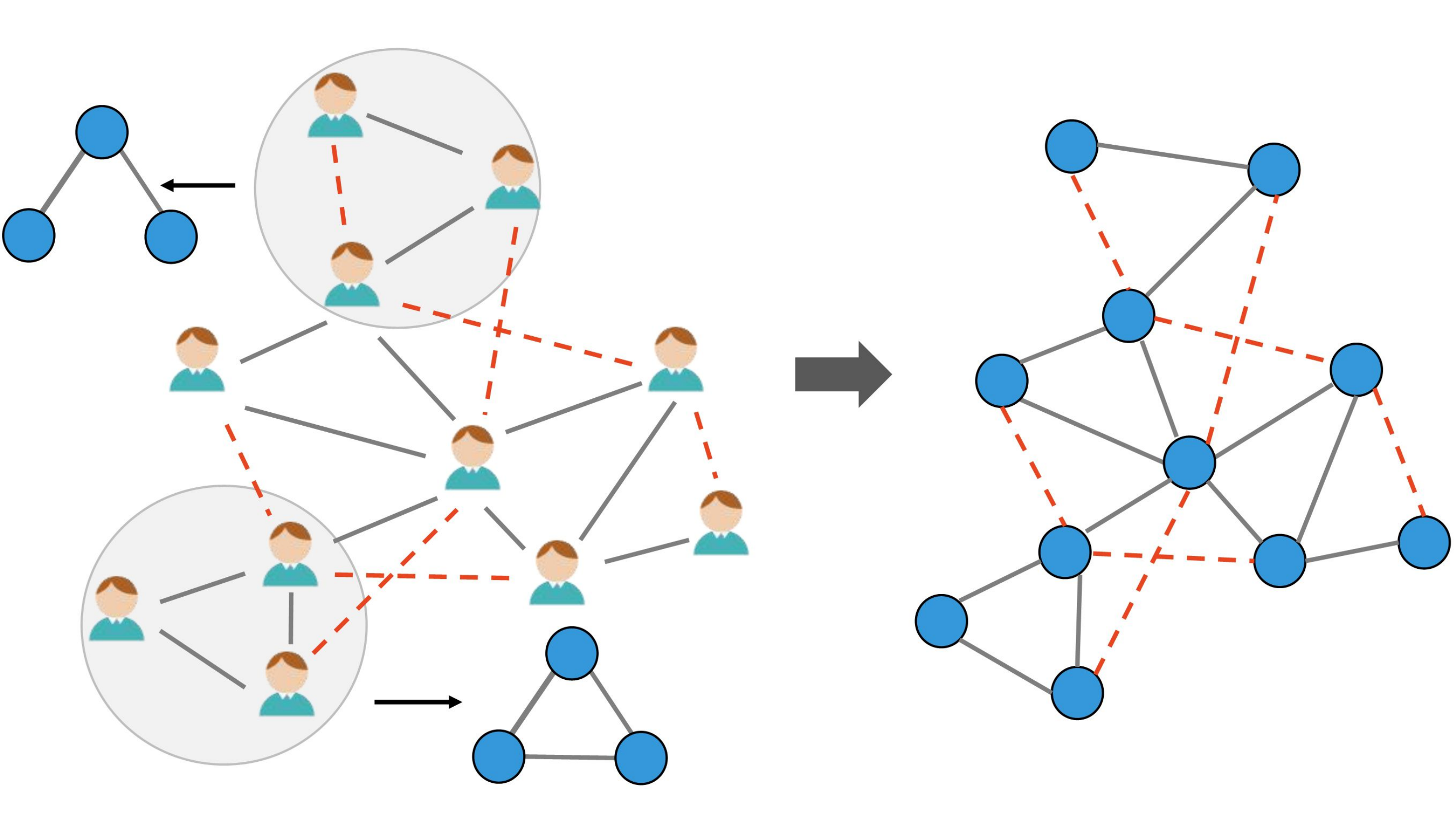}
  \caption{Triangle closures in academic social networks. The left figure shows the academic social network in reality, and the arrow represents the types of motifs that exist in the network. The right figure shows the topology of the network. The blue nodes represent scholars, the gray edges represent the cooperative relationship between them, and the red dashed lines represent possible links.}\label{fig:motif}
\end{figure}

An academic collaboration network is generally constructed by analyzing groups of scholars who have produced research publications together. This network is constructed based on authorship and co-authorship relationships. In the constructed network, nodes represent scholars and edges represent the collaboration relations between certain scholars. The weights of edges are generally qualified by the number of publications two scholars published, which reflects the collaboration intensity between two scholars to some extent. It has been discovered that triangle closures are common in  academic social networks. These can be represented by network motifs (shown in Fig.\ref{fig:motif})~\cite{xu2020multivariate}. Network motifs are induced subgraphs that appear more frequently in real-world networks, which generally represent certain social patterns that have real meaning~\cite{zhang2020graph}.

By exploring the concept of network motifs, this work uses familiarity~\cite{yu2020optimize} based on one’s network structure and then proposes a team recognition method. We first calculate the lengths of possible paths between two nodes in the network. Then we calculate the local density and distinguishable distance between each node to draw a decision figure. Nodes with greater local density and a greater distinguishable distance value will be chosen as central nodes for each team. The remaining nodes are linked with a particular central node based on their shortest distances to the central node, local density, and familiarity. The contributions of this paper are summarized below:
\begin{itemize}
\item  \textbf{Collaborative team recognition:} We propose MOTO (faMiliarity-based cOllaborative Team recOgnition algorithm) to recognize collaborative teams in academia. The proposed approach employs network motifs to qualify familiarity among scholars and recognize teams with a local density as well as a distinguishable distance between nodes.
\item  \textbf{Higher-order familiarity qualification:} We employ the qualification metric of higher-order familiarity among scholars. As the structural property of a certain node, the number of motifs can reflect the collaboration familiarity among scholars. Based on this metric, an academic team is recognized more precisely.
\item  \textbf{Real-world data verification:} We use a real-world data set to recognize collaborative teams. Microsoft Academic Graph Computer Science data set is employed in the experiments. The experimental results show that our proposed method can recognize collaborative teams effectively. These teams are then analyzed in more detail.
\end{itemize}

The remainder of this paper is organized as follows. Section~\ref{sec:2} introduces related work, including network structure and team recognition. Section~\ref{sec:3} introduces preliminaries including the definition of pairwise familiarity, higher-order familiarity, and network motif. Section~\ref{sec:4} illustrates the details of MOTO algorithm. Section~\ref{sec:5} analyzes experimental results. Finally, Section~\ref{sec:6} concludes the paper.

\section{Related Work}\label{sec:2}
\subsection{Network Structure}
The definition of network motifs and the algorithm for mining them was first provided by Milo et al.~\cite{milo2002network}. They were attempting to identify patterns in complex networks and found frequent occurrences of subgraph connection patterns that would not be found in equivalent numbers in random networks.  These recurring, significant patters of interconnections were given the title of network motifs and Milo and his colleagues found examples in biochemistry, neuorbiology, ecology and engineering networks. Since that time, researchers have continued to find applications for network motifs in multiple research areas~\cite{xia2019survey}.

The discovery and counting of network motifs has become an important research area. Ahmed et al.~\cite{ahmed2015efficient} proposed a fast and efficient parallel counting method for three-point and four-point subgraph patterns. The method can calculate the accurate number of subgraph patterns and significantly reduce the calculation time.  Ma et al.~\cite{ma2019linc} explored solutions to the problem of counting motifs on uncertain graphs and proposed two algorithms named Possible Graph Sampling (PGS) and Linking and Counting (LINC). PGS samples some possible worlds from the graph and then runs a deterministic modal counting algorithm on each possible world. LINC is an improvement of PGS; it can effectively calculate the difference in the motif count of two different possible worlds. Different algorithms for the discovery of network motifs have also been studied. Yu et al.~\cite{yu2020motif} classified and summarized the discovery algorithms of network motifs and compared the running time of different algorithms. They also discussed the application of these algorithms in various scenarios.

Many studies use network motifs to analyze the characteristics of different types of networks. Milo et al.~\cite{milo2004superfamilies} analyzed the distribution of triangle motifs and four-order motifs in different networks, and classified networks in different fields according to the statistical importance in the distribution curve of the number of motifs. They identified that the statistical importance of the triangle motif in social networks is significantly higher than in others networks. Zhao et al.~\cite{zhao2019ranking} studied network motifs in social networks and proposed a method called Motif-based PageRank (MPR), which considers first-order and higher-order relations for user ranking in social networks. They computed the motif-based adjacency matrix and combined it with the edge-based adjacency matrix to re-weigh the links between users. They also studied the performance of other types of motifs. Paranjape et al.~\cite{paranjape2017motifs} defined and studied motifs in time series networks.

Some research has combined an application of clustering and motifs with large-scale networks. Benson et al.~\cite{benson2016higher} developed a general framework based on network clustering of high-order motifs. They showed that there are rich high-order motifs in large-scale networks, such as the information dissemination unit of neural networks and the network hub structure of traffic networks. To solve the dynamic local motif clustering problem, Fu et al.~\cite{fu2020local} proposed a model called Local Motif Clustering on Time-Evolving Graphs (L-MEGA). L-MEGA mainly used edge filtering, motif push operation, and incremental sweep cut to track the temporal evolution of the local motif cluster.

More recent studies have continued to systematically review network motifs. Yu et al.~\cite{yu2019motifs} summarized the definition and related concepts of network motifs. They analyzed network motifs in biological networks, social networks, academic networks, and infrastructure networks. They provided insights into motif discovery, motif technology, motif clustering methods, and network motif applications in different fields. Xia et al.~\cite{xia2019survey} classified network motif measures into structural measures and statistical measures according to the calculation method of the measurement indicators. They analyzed the application of these measures in the discovery, counting, analysis, and clustering of the network motif.

These studies all support the notion that network motifs can reveal the basic structure of most networks and play an important role in various network applications. However, most of the existing research ignores the influence of the network structure and the familiarity between team members and its influence on recognition. To address these issues, we comprehensively consider these factors and use the existing familiarity metric to quantitatively describe the familiarity between scholars. We propose MOTO based on this metric, which makes the identified team more cohesive and lowers the cost of team communication.

\subsection{Team Recognition}
Academic team recognition algorithms have evolved corresponding to changes in the nature of academic teams over time. Before the large-scale development of the Internet and social networks, academic teams were the same as scientific research institutions, referring to scholars engaged in scientific research in the same institution. Traditional academic team recognition methods have used artificial methods such as questionnaire surveys; these methods have low efficiency, are high time consuming and costly, and are limited by the available samples produced. With the rapid development of social networks, scholars have been able to cooperate remotely and a large number of interagency and interdisciplinary academic teams have emerged. However, the concept of what an academic team actually is, is not settled. For example, some studies regard the co-author of a paper as a member of a research team and have used this definition to explore the macro issues of team science~\cite{chen2017building,molontay2019two}. Some researchers regard a team with two to ten scientists as a scientific team and a team with more than ten as a large team~\cite{cooke2015enhancing}. Some researchers have provided their own definition based on the classic definition. Some studies use visual tools to show networks and combine cliques 
 to identify academic teams~\cite{shi2016survey}. However, in reality, the members of these identified teams may not directly collaborate. Calero et al.~\cite{calero2006identify} proposed a new bibliometric method to identify research teams in specific research fields by combining bibliometric methods and network analysis. Yu et al.~\cite{yu2017team} proposed a team recognition method called TRAC based on the Collaboration Intensity Index. The method uses a top-down approach to delete edges with a cooperation intensity less than a threshold, and finally, uses the derived small connected networks to define academic teams.

Community detection algorithms can be used to discover the community structure in networks~\cite{xia2021gl}. However, when classic community detection algorithms that deal only with the structure of social networks or detect communities using only node attributes e.g. age, gender and interests are used in isolation, the results may be limited~\cite{chunaev2020community}. Team recognition tasks are more fine-grained, the team members have different attributes and this makes relationships complicated. Therefore, the following research improves the community detection algorithm to identify academic teams. Savi{\'c} et al.~\cite{savic2016community} proposed a method for community detection in research collaboration networks. They set frequent collaborators as the core of the research team and determined them through w-core. W-core is a graph traversal algorithm, which mainly assigns each node so that the two nodes from the same w-core have the same label, while the two nodes from different w-core have different labels. Villarreal et al.~\cite{villarreal2016local} proposed a clustering algorithm for cooperative scientific networks, where attempts are made to cluster on both sides of a bipartite graph in order to obtain the cluster of authors and articles. The proposed method not only detects research teams, but also describes and visualizes the detected teams.

Yu et al.~\cite{Yu2018themethod} 
use a slightly different approach by defining an academic team as an academic cooperative team composed of leaders, core team members and non-core members. Their research teams identification method identifies team leaders based on the centrality measure and uses 2-clique to identify core members. However, their approach does not take into consideration the degree of relationship between team members, such as the closeness and familiarity of the connection, which makes it difficult to identify academic teams efficiently.

Most of existing community detection methods are complicated and have high computational costs. Therefore, in this work, we propose a team recognition method by exploring cluster centers. Compared to the community detection approaches, the design of our method is straightforward, which can recognize clusters regardless of their shape and the dimensionality of the space in which they are embedded. Moreover, academic teams are generally with “core+extended” structure\cite{xia2021chief}. Our proposed method can better recognize teams with such structure.

\section{Preliminaries}\label{sec:3}
In keeping with the objectives of this paper, we now in this section provide a definition of familiarity, which is used to measure the overall familiarity between scholars and other team members. We also provide a more formal definition of the concept of network motifs.

\subsection{Familiarity}
Yu et al.~\cite{yu2020optimize} first proposed the definition of familiarity. They divided it into Pairwise familiarity and Higher-order familiarity. The following is the specific calculation formula.

\subsubsection{Pairwise Familiarity}
Pairing familiarity refers to the number of team members who have established a cooperative relationship with the scholar, i.e., there are edges in the cooperative network. The formula is shown in Eq.\eqref{eq:pf}.
\begin{equation}\label{eq:pf}
\|F\|_{1}(i, T)=\sum_{j \in T, j \neq i}{PairwiseCol}_{i j}
\end{equation}
where T refers to the team. When there is an edge between $i$ and $j$, ${PairwiseCol}_{i j} = 1$, otherwise ${PairwiseCol}_{i j} = 0$. For scholars outside the team T, the more people they have worked with in the team, the more familiar the scholar is with the team. The communication cost of cooperation is even lower.

\subsubsection{Higher-order Familiarity}
Higher-order familiarity refers to the number of team members who have established a relatively stable cooperative relationship with the scholar. Relatively stable means they have established a triangle motif. The formula is shown in Eq.\eqref{eq:hof}.
\begin{equation}\label{eq:hof}
\|F\|_{n}(i, T)=\sum_{j \in T, j \neq i}{ MultiCol}_{i j}
\end{equation}
where ${MultiCol}_{i j} = 1$ means that $i$ and $j$ appear in at least one triangle motif, ${MultiCol}_{i j} = 0$ means that $i$ and $j$ never formed a triangle motif. $\|F\|_{n}(i, T)$ indicates that the number of members in team $T$ who form a triangle motif with $i$. The higher of $\|F\|_{n}$, the more familiar $i$ is with team $T$.

\subsection{Network Motif}
Milo et al.~\cite{milo2002network} first proposed the definition of network motif. They proposed that network motifs are interconnections patterns of the subgraph that repeatedly appears in the original network, which appears more frequently than in the similar random network. The distribution of node degree in random networks and real networks should be consistent.

Let $G=\{V,E\}$ be a network, where $V$ is the node set, $E$ is the edge set. $G_k \subset G$ means the subgraph of $G$ whose size is $k$. Given a network $G$, a set of parameters $\{P, U, D, N\}$ and a set of $N$ similar random networks. The network motif is defined as an induced subgraph appearing in the real network that meets the following three conditions:

  \begin{equation}
  p(\bar f_{rand}(G_k)>f_{real}(G_k)) \leq P  \nonumber
  \end{equation}
  
  \begin{equation}
  f_{real}(G_k) \le U  \nonumber
  \end{equation}

  \begin{equation}
  f_{real}(G_k) - \bar f_{rand}(G_k) > D \times \bar f_{rand}(G_k)  \nonumber
  \end{equation}
where $f_{real}(G_k)$ is the occurrence of the subgraph in the real network, $\bar f_{rand}(G_k)$ is the average occurrence of the subgraph in all random networks. $P$ is the probability threshold determined by $N$ similar random networks. The first condition is to ensure the motif did appear with much higher frequency in real-world network comparing to random network. $U$ is the unique cutoff value of the frequency of network motif in the real network. The second condition is to limit the appearing frequency of motif that appears in real-world network. In the third condition, $D$ is the minimum difference cutoff ratio to ensure the minimum difference between $f_{real}(G_k)$ and  $\bar f_{rand}(G_k)$. According to the experience, the parameters $\{P, U, D, N\}$ are generally set as $\{0.01, 4, 0.1, 1000\}$.

\begin{figure}[tb]
  \centering
  \includegraphics[width=0.9\linewidth]{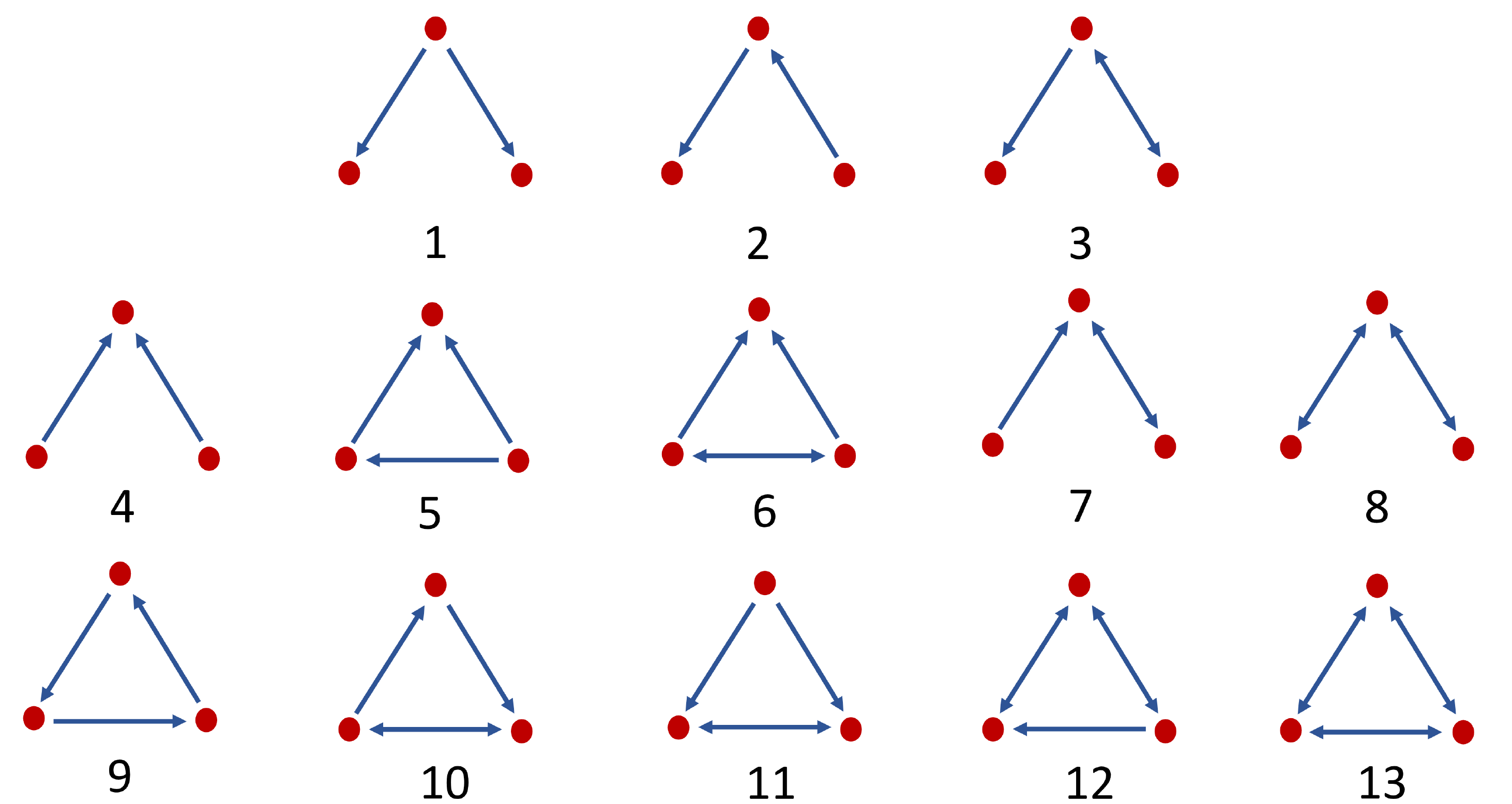}
  \caption{All possible directed 3-motifs.}\label{fig:3-motifs}
\end{figure}

Fig.~\ref{fig:3-motifs} shows all possible directed 3-motifs. Here we only list motifs with two or more edges to make sure that all the listed motifs are connected. At present, many studies have identified motifs with distinctive characteristics in different types of networks. For example, the triangle fully connected motif appears more frequently than other motifs in social networks~\cite{milo2004superfamilies}. These triangle fully connected motifs are demonstrated in subfigures 9 to 13 of Fig.~\ref{fig:3-motifs}. Such findings have motivated researchers to take advantage of the characteristics of motifs in relevant research. In particular, collaboration relationships are recognized as two-way edges. Consequently, in this work, motifs are regarded as being undirected in line with undirected collaboration networks.

\begin{figure*}[!tb]
  \centering
  \includegraphics[width=0.8\linewidth]{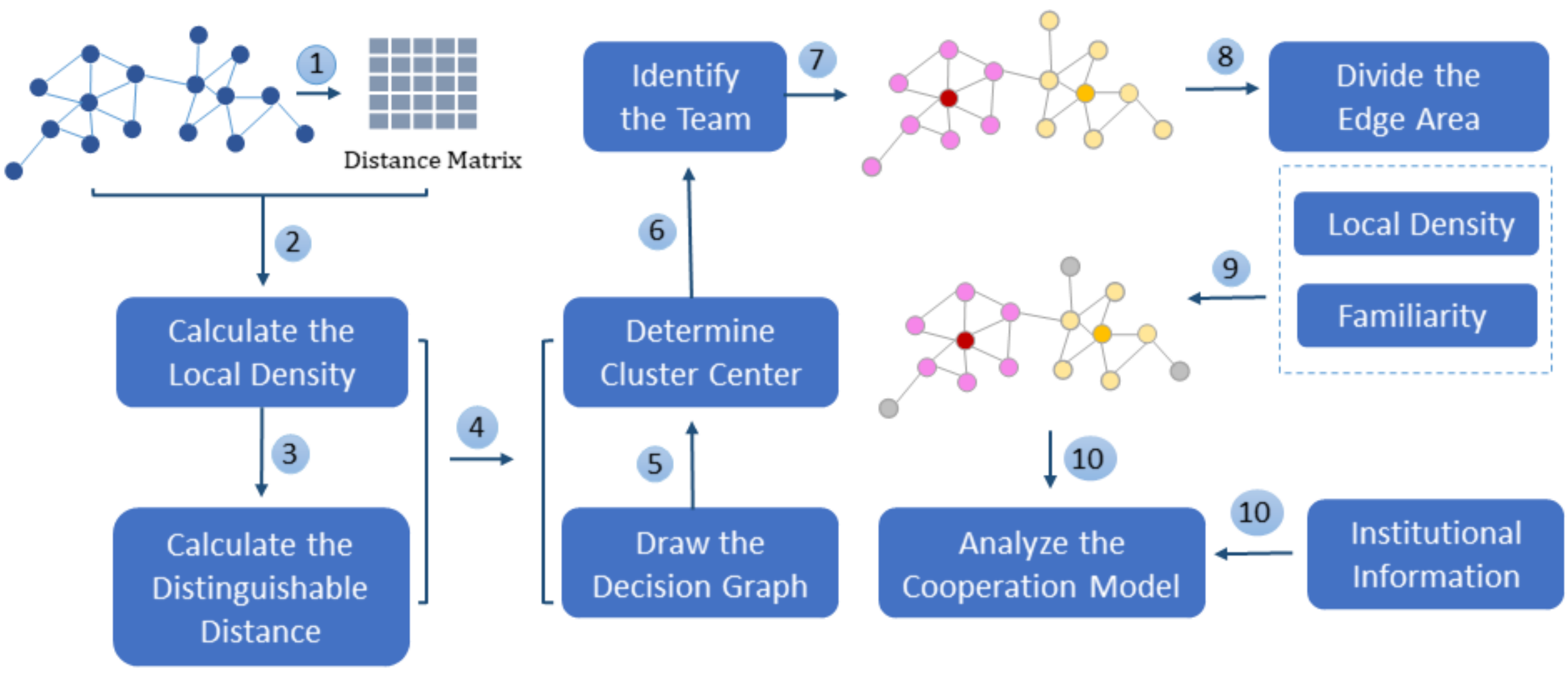}
  \caption{The flowchart of MOTO.}\label{fig:algorithm-progress}
\end{figure*}

\section{The Design of MOTO}\label{sec:4}
In an academic team, the familiarity between members is an important feature of the team. Some studies have shown that people are more inclined to cooperate with familiar people, and team members are more familiar with their team than others~\cite{gu2019exploring}. In this section, we propose MOTO, which is based on the CFSFDP algorithm proposed by Alex et al.~\cite{rodriguez2014clustering}.

The main steps of MOTO are shown in Fig.\ref{fig:algorithm-progress}. Firstly, we add weights to the edges in the academic collaboration network, and then calculate the shortest distance matrix between any two nodes. Next, calculate the local density of each node and the minimum distance between each node and all nodes with higher local density. Determine the cluster center node and the number of clusters based on the local density of each node and the maximum distance from the high-density node. After that, assign other nodes except the center nodes to the nearest cluster center node so as to complete the preliminary team recognition. Then divide the edge area of each team based on the familiarity and determine the threshold of the local density and team familiarity of the team members. Filter the team members according to the two thresholds to identify the academic team. Finally, we divide these academic teams with academic institutions and get the academic teams within the academic institutions. The following sections will describe major steps of MOTO in detail.

\subsection{Calculation of Node Pair Distance}
In step one, we calculate the distance between all scholars in the network. $G$ is an undirected weighted graph. The weight of an edge is the cooperation distance between two nodes, which is shown in Eq.\eqref{eq:dij}.
\begin{equation}\label{eq:dij}
d_{i j}=1-\frac{\left|P_{i} \cap P_{j}\right|}{\left|P_{i} \cup P_{j}\right|}=1-\frac{\cot _{i j}}{p n_{i}+p n_{i}-\cot _{i j}}
\end{equation}
where $P_i$ and $P_j$ represent the paper set of scholar $i$ and $j$, respectively. $|P_{i} \cap P_{j}|$ is the number of papers co-authored by scholars $i$ and $j$.  $|P_{i} \cup P_{j}|$ is the number of non-repeated papers written by the two scholars. The minimum value of $d_{ij}$ is 0, and the maximum value is 1. In order to improve the efficiency of calculation, we simplify it. We represent $|P_{i} \cap P_{j}|$ as ${cot}_{ij}$, which refers to the times of cooperation between scholar $i$ and $j$. ${pn}_i$ and ${pn}_j$ represent the number of papers by scholar $i$ and $j$, respectively.

In the subsequent clustering step, we need to calculate the distance between any two nodes, which is the sum of the distance between two nodes of the shortest path. It is represented as
\begin{equation}\label{eq:dis}
dis(v_i,v_j)=shortestPathLength(v_i,v_j)
\end{equation}

We choose the Dijkstra algorithm to calculate the distance. The Dijkstra algorithm is the shortest path algorithm from one node to the other nodes. It is applicable to both directed and undirected graphs, and it requires the weight to be non-negative. Due to the large scale of the academic collaboration network, the exploration range value of the shortest path can be set in the calculation process. This can reduce the complexity whilst calculating the distance required for clustering.

\subsection{Calculation of Local Density and Distinguishable Distance}
In step two, we calculate the local density $\rho$ of each node within the cutoff distance $d_c$. The $\rho$ of a scholar in the network measures the density of scholars who have a certain degree of close cooperation with the scholar. $d_c$ is the only hyper-parameter in the algorithm, which represents the range of $\rho$. For each node $i \in V$ in $G$, $\rho_i$ refers to the number of other nodes in the network whose distance from node $v_i$ does not exceed the range of $d_c$ except for node $v_i$. It can be calculated using the following equation:

\begin{equation}
 \rho_{v_i}=\sum_{v_j \in V, v_j \neq v_i} \chi\left(dis\left(v_i, v_j\right)-d_c\right)
\end{equation}

\begin{equation}
\chi(x)=\left\{
\begin{array}{ll}
1, & x<0 \\
0, & x \geq 0
\end{array}
\right.
\end{equation}
where the value of function $\chi(x)$ is 1 when the distance between $v_i$ and $v_j$ is less than $d_c$, otherwise the value is 0. It should be noted that we believe that the $\rho$ of the cluster center node is very high. This is specifically reflected in the fact that the scholar keeps collaboration with more people in the team, rather than that the scholar is the leader of the team. Then we sort all nodes in descending order according to their $\rho$. The high-density node set of node $v_i$ is $VP_{v_{i}}=\left\{v_{j} \mid \rho_{v_{j}}>\rho_{v_{i}}\right\}$.

Next, we calculate the shortest distance between each node and the high-density node, i.e., the distinguishable distance $\delta$. We use the distance between two nodes to distinguish between two teams so if we assume that a node is a cluster center node, the node with greater $\rho$ than this node is either the center of another team or the node closer to the center in the same team. In other words, in the cluster where a central node is located, there should be no nodes with higher $\rho$ than it. Therefore, when determining the cluster center, in order to ensure that the distance between the clusters is significant, the cluster center should be further away from all the higher density nodes than those within its cluster so that the two clusters will not merge into one cluster. The $\delta$ of node $v_i$ is the minimum distance between $v_i$ and $VP_{v_{i}}$, defined as

\begin{equation}
\delta_{v_i}=\left\{\begin{array}{c}
\min\limits _{v_{j} \in VP} \operatorname{dis}\left(v_{i}, v_{j}\right), \rho_{v_{i}} \neq \max _{-} \rho \\
\max\limits _{v_{j} \in V, v_{j} \neq v_{i}} \operatorname{dis}\left(v_{i}, v_{j}\right), \rho_{v_{i}}=\max _{-} \rho
\end{array}\right.
\end{equation}
where $\max _{-} \rho$ is the maximum $\rho$ of all nodes. For the node with the highest $\rho$, its distinguishable distance is the maximum distance from any other node.

\subsection{Determine Cluster Center}
In step three, we use the local density and distinguishable distance of each node to draw the cluster center decision graph, as shown in Fig.\ref{fig:devision}. The horizontal axis represents the local density, and the vertical axis implies the distinguishable distance. The decision graph is divided into four areas, and the nodes in each area have different characteristics:
\begin{figure}[htb]
  \centering
  \includegraphics[width=6cm]{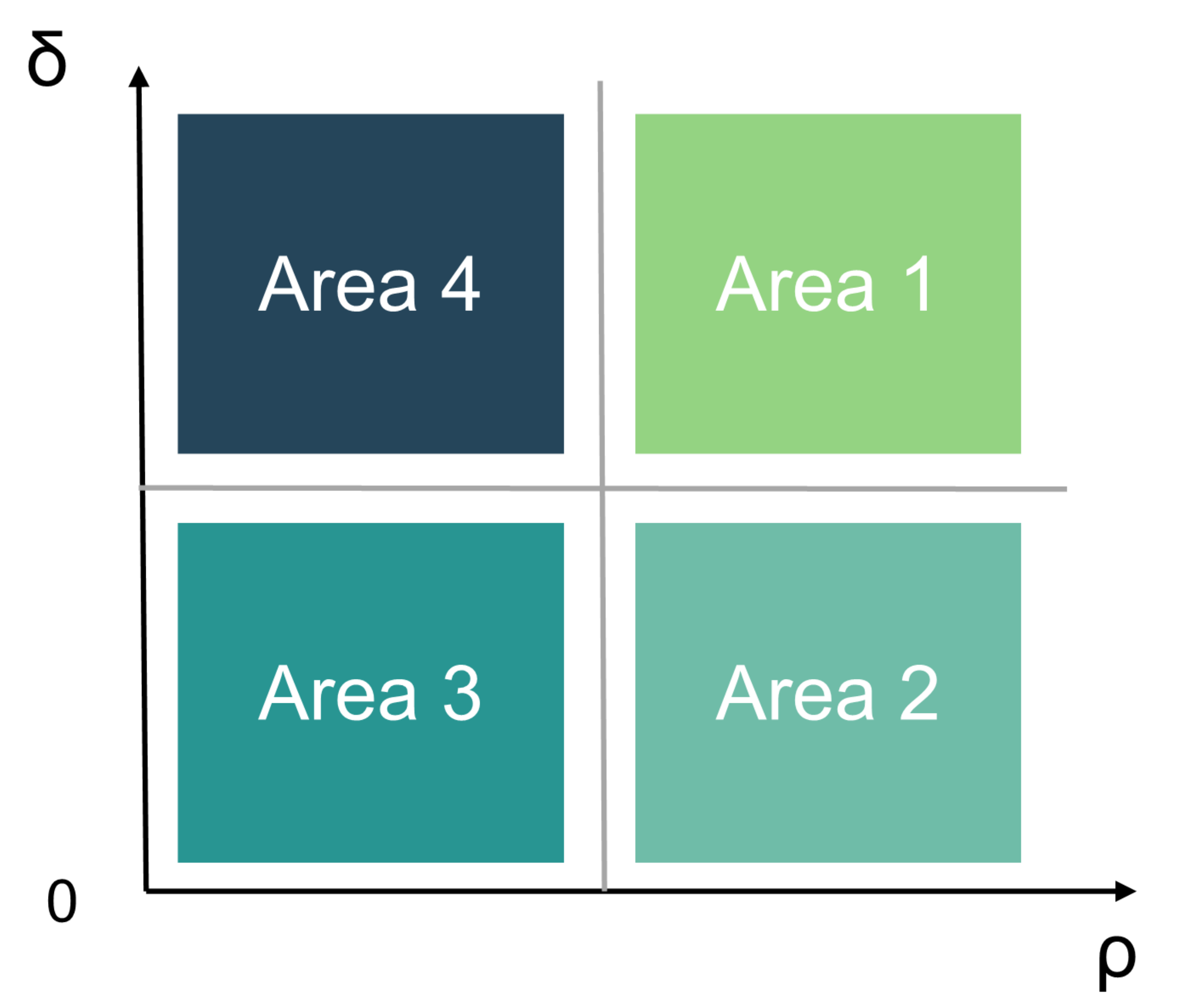}
  \caption{The regional division of decision graph.}\label{fig:devision}
\end{figure}
\begin{itemize}
  \item  \textbf{Area one:} The nodes have high $\rho$ and high $\delta$, which accords with the characteristics of high $\rho$ of cluster center nodes and are some distance from other possible cluster center nodes. So the nodes in this area are cluster center nodes. This area can be clearly distinguished from other areas.
  \item  \textbf{Area two:} The nodes have high $\rho$ but low $\delta$. These nodes’ cutoff distances contain nodes with higher $\rho$, which are nodes close to the center but not the center in the team. The specific roles of these nodes in the team need to be analyzed in combination with experimental results.
  \item  \textbf{Area three:} The nodes have low $\rho$ and low $\delta$. These points are located in relatively sparse locations in the network, and comparatively further away from all nodes with high $\rho$. They may be located in the middle of different circles, or the collaborators are relatively scattered. Each specific situation should be analyzed based on the experimental results.
  \item  \textbf{Area four:} The nodes have low $\rho$ but high $\delta$, which means that they are relatively isolated nodes. Such scholars have few collaborators.
\end{itemize}

Due to the larger scale of the academic collaboration network, it is still necessary to apply a more intuitive judgment method when observing the decision graph to get the  local density of the cluster center area and the boundary of the distinguishable distance. We calculate the product of $\rho$ and $\delta$ of each node, i.e., $\gamma_{v_i} = \rho_{v_i} \times \delta_{v_i}$. Therefore, according to $\gamma$ and the decision graph, we can get a set $Center=\{c\}$ consisting of nodes with relatively high $\rho$ and $\delta$.

\subsection{Team Recognition}
In this step, after having identified the cluster center nodes in step three, we divide the entire network into $\mid Center \mid$ clusters, i.e., academic teams. For each non-cluster center node $v_i$, we calculate the distance $dis(v_i,c)$ between the node and each central node, and place the node into the cluster where the nearest cluster center node is located.

Next, we identify the set of eligible nodes in each cluster. Academic team members are closely connected with the team and maintain a certain degree of familiarity with the team members but are relatively sparsely connected with other teams . That is, the cooperative behavior of a team member should have the familiarity and closeness of connecting with other members of the team. Team familiarity means the sense of participation in the team, i.e., the member has direct collaboration or high-level collaboration relationship with the team members. Closeness measures the local density degree of members in the collaboration network. A node with higher closeness is more probably to be recognized as central node. Therefore, Closeness is employed to recognized cluster centers and familiarity is employed to recognize other team members corresponding to centers. In summary, academic team members should meet the following two conditions:

1) \textbf{Closeness:} $\rho$ is higher than the threshold $\rho'$.

2) \textbf{Familiarity:} the team familiarity is higher than the threshold $\|F\|'_{n}$.

Determining the values of $\rho'$ and $\|F\|'_{n}$ is an important step in filtering. Because the team size and the closeness of cooperation are different, it is necessary to determine each team's threshold based on the characteristics of each team. First, divide the edge area of the team based on the condition that there are member nodes of different teams within the neighborhood of the node's cutoff distance, then the edge area $border(T)$ of the team $T$ is represented as:
\begin{equation}
\text {border}(T)=\left\{v_{i} \mid \exists \operatorname{dis}\left(v_{i}, v_{j}\right)<d_{c}, v_{i} \in T, v_{j} \notin T\right\}
\end{equation}

\begin{algorithm}
  \caption{MOTO}  
  \textbf{Input:}~$G=(V,E,W)$, which is the academic collaboration graph with edge weight (collaborative distance between scholars), cutoff distance $d_c$;\\
  \textbf{Output:}~academic team list ${T_c}$;
  \begin{algorithmic}[1]
    \State $\rho_{v_i}$,$\delta_{v_i}$ = Calculation of Local density $\rho_{v_i}$ and distinguishable distance  $\delta_{v_i}$ (G, $d_c$)
    \State $Cluster$ = Division of Academic Teams Algorithm (G, $\rho_{v_i}$, $\delta_{v_i}$)
    \For {each $T$ of $Cluster$}
    \For {$v_i$ in $V$}
    
    \State calculate $\|F\|_{n}(v_i, T)$;
    \EndFor
    
    \State ${border}(T)=\left\{v_{i} \mid \exists \operatorname{dis}\left(v_{i}, v_{j}\right)<d_{c}, v_{i} \in T, v_{j} \notin T\right\}$
    \State $\rho'=\max \rho,\rho\in border(T)$
    \State $\|F\|'_{n}=\max\|F\|_{n},\|F\|_{n} \in border(T)$
    \State $T_{c}=\left\{\left.v_{i}\left|\rho_{v_{i}} \geq \rho^{\prime},\right||F|\right|_{n}\left(v_{i}, T\right) \geq|| F||_{n^{\prime}}^{\prime} v_{i} \in T\right\}$
    \EndFor \\
    \Return $T_c$
  \end{algorithmic}
\label{alg:TR}
\end{algorithm}

\begin{algorithm}
  \caption{Calculation of Local density $\rho_{v_i}$ and distinguishable distance  $\delta_{v_i}$}
    
    \textbf{Input:}~$G=(V,E,W)$, which is the academic collaboration graph with edge weight (collaborative distance between scholars), cutoff distance $d_c$;\\
    \textbf{Output:}~local density $\rho_{v_i}$, distinguishable distance  $\delta_{v_i}$ ;
  \begin{algorithmic}[1]
    \For {$pair(v_i, v_j)$ in $V$}
      \State calculate $dis(v_i, v_j)$;
    \EndFor
    
    \For {$v_i$ in $V$}
      \State $\rho_{v_i}=\sum_{v_j \in V, v_j \neq v_i} \chi\left(dis\left(v_i, v_j\right)-d_c\right)$;
    \EndFor
    \For {$v_i$ in $V$}
  
      \State $VP_{v_{i}}=\left\{v_{j} \mid \rho_{v_{j}}>\rho_{v_{i}}\right\}$;
    \EndFor
    \For {$v_i$ in $V$}   
  
      \If {$\rho_{v_{i}} \neq \max _{-} \rho$}
        \State $\delta_{v_i}=\min\limits _{v_{j} \in VP} \operatorname{dis}\left(v_{i}, v_{j}\right)$;
      \Else
        \State $\delta_{v_i}=\max\limits _{v_{j} \in V, v_{j} \neq v_{i}} \operatorname{dis}\left(v_{i}, v_{j}\right)$
      \EndIf
  
    \EndFor \\
  \Return  $\rho_{v_i}$,$\delta_{v_i}$
\end{algorithmic}
\label{alg:calculate}
\end{algorithm}

\begin{algorithm}
\caption{Division of Academic Teams Algorithm}

 \textbf{Input:}~$G=(V,E,W)$, which is the academic collaboration graph with edge weight (collaborative distance between scholars);\\
\textbf{Output:}~ cluster set $Cluster$ ;
\begin{algorithmic}[1]  
  
    \For {$v_i$ in $V$}
      \State $\gamma_{v_i} = normalized(\rho_{v_i} \times \delta_{v_i})$
    \EndFor
    \State Select the cluster center node set $Center$
  
    \For {$v_i$ in $V$}
    \State $Cluster_{v_i} = \mathop{\arg\min}\limits_ {c_j \in Center}dis(v_i,c_j)$
    \EndFor \\
  \Return $Cluster$
  \end{algorithmic}
\label{alg:cluster}
\end{algorithm}

The nodes in the team edge area are the nodes where the cooperation between the team and other teams is not significant. Additionally the local density of these nodes is not strong enough to be a cluster center, nor are they isolated scholars. Team familiarity is similar to the situation of these nodes. Therefore we choose the maximum local density and team familiarity of the node in the edge region as the thresholds for $\rho'$ and $\|F\|'_{n}$.

The next step is to filter all members of the team according to $\rho'$ and $\|F\|'_{n}$. Nodes with local density and familiarity above the threshold are identified as team members. The team edge nodes are the boundary part of multiple teams or a single team, which can be successfully identified by the above filtering methods. The set of teams obtained is expressed as $T_c$:
\begin{equation}
T_{c}=\left\{\left.v_{i}\left|\rho_{v_{i}} \geq \rho^{\prime},\right||F|\right|_{n}\left(v_{i}, T\right) \geq|| F||_{n^{\prime}}^{\prime} v_{i} \in T\right\}
\end{equation}

Finally, the division of nodes according to institutional attributes and the division of academic teams obtained by clustering are shown in the network.  An academic team $TI_{c}^{i}$ in $Institution_k$ is represented as:
\begin{equation}
TI_{c}^{i}=\left\{v_{j}\left|v_{j} \in T_{c},\right| v_{j} \in \text {Institution}_{k}\right\}
\end{equation}

These are the main processes of MOTO, and the specific pseudo-codes are shown in Algorithm~\ref{alg:TR}. The parameter setting varies according to the different parameter values of the academic collaboration network.

\section{Experiments and Results}\label{sec:5}
The focus of this section is to introduce the dataset used in the experimentation, the data preprocessing process, the network statistics overview and the experimental settings. In order to evaluate our proposed algorithm, we also introduce a number of baseline methods and analyze the experimental results.

\subsection{Dataset Collection and Data Preprocessing}
\begin{figure}[htb]
  \centering
  \includegraphics[width=0.8\linewidth]{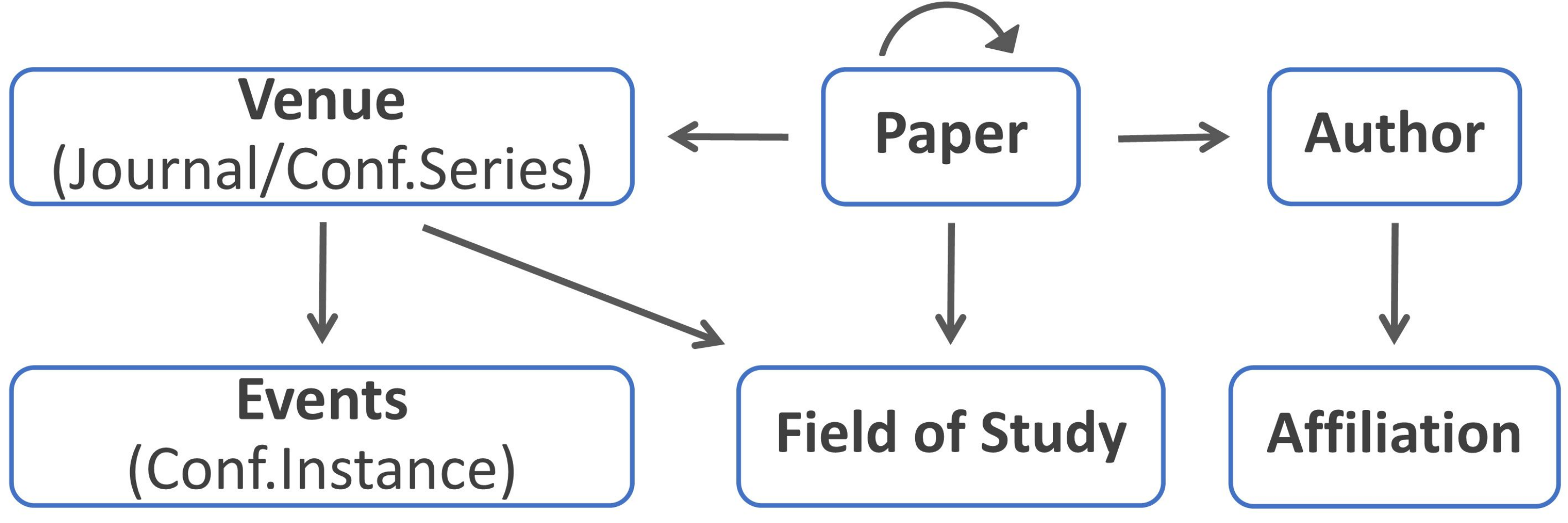}
  \caption{The relationship of entities in MAG data set.}\label{fig:mag}
\end{figure}

We conducted extensive experiments using data from Microsoft Academic Graph (MAG)\footnote{https://www.microsoft.com/en-us/research/project/microsoft-academic-graph/}. It is an open academic dataset that contains more than 200 million scientific research literature publication records and citation relationships between the literature since 1800 ~\cite{sinha2015overview}. MAG includes six entities: publications, authors, institutions, journals, conferences, and fields of study. The relationships between entities are shown in Fig.\ref{fig:mag}.

For the MAG dataset, we perform the following processing operations to obtain the experimental data.

\begin{enumerate}
\item Firstly, we extract all papers in computer-related fields. We select papers in these fields from 2006 to 2017 as experimental data. MAG contains 34 sub-areas in Computer Science. According to the attribute fieldOfStudy of the paper, we extract 12,923,247 papers related to the above 35 fields. We then extract papers in the desired year. Yu et al.~\cite{yu2017team} showed that if two scholars collaborated and did not cooperate again in the next four years, they would not cooperate again. So the experiment focused on selecting periods of four years. Due to the evolutionary behavior of the team, each period is separated by two years. After filtering papers with missing information, we extracted 1,066,628 papers from 2006 to 2009, 1,258,318 papers from 2008 to 2011, 1,477,560 papers from 2010 to 2013, 1,602,827 papers from 2012 to 2015, and 2,827,671 papers from 2014 to 2017.

\item Secondly, we filtered the authors. Although students are also involved in scientific research, most of them will end their academic lives within 5 years and do not constitute the backbone of the academic team. Therefore, we select scholars with an academic life of 5 years or more as the research objects. Finally, we obtained 291,188 scholars who meet the above requirements in 12 years. Then, we constructed the academic collaboration network by the cooperative relationship between these scholars.
\end{enumerate}

After constructing the academic collaboration network, we found that the network contained many connected pieces and nodes. To more easily control the experiment, we selected the largest connected piece in the collaboration network in each period to use in subsequent experiments. The profile of the academic collaboration network is shown in Tab.\ref{tab:profile}. As the network grew in size over time, the average number of collaborations between scholars and the average degree of nodes increased. This indicates that the number of each scholar's collaborators also increased.

\begin{table}[htbp]
  \centering
  \caption{The profile of the academic collaboration network.}
  \resizebox{0.48\textwidth}{!}{
    \begin{tabular}{cccc}
      \toprule
      \textbf{Properties} & \textbf{Nodes} & \textbf{Edges} & \textbf{Avg Co-times}   \\
      \midrule
      \textbf{2006-2009} & 105,721 & 298,768 & 2.163   \\
      \textbf{2008-2011} & 140,241 & 429,253 & 2.092  \\
      \textbf{2010-2013} & 167,535 & 544,986 & 2.033  \\
      \textbf{2012-2015} & 179,773 & 602,301 & 2       \\
      \textbf{2014-2017} & 197,001 & 832,248 & 2.449   \\
      
      \midrule
        \textbf{Properties} &  \textbf{Avg degree} & \textbf{Triangles} & \textbf{CCF} \\
        \midrule
      \textbf{2006-2009}   & 5.652 & 416,454 & 0.38 \\
      \textbf{2008-2011}  & 6.121 & 648,055 & 0.391 \\
      \textbf{2010-2013}  & 6.505 & 1,368,684 & 0.394 \\
      \textbf{2012-2015}   & 6.701 & 1,146,961 & 0.397 \\
      \textbf{2014-2017}   & 8.45  & 1,843,896 & 0.385 \\
      \bottomrule
    \end{tabular}%

%
  }
  \label{tab:profile}%
\end{table}%

An important qualification that we needed to make was that the institution a scholar belonged to, was not necessarily fixed over time i.e. they may not belong to the same institution across  different periods. So we extracted the author's institution and removed duplicates to get the author's full institutional attributes. If a scholar had more than one institution in a certain period, the scholar was considered to belong to both those institutions at the same time. The institutional attributes of scholars are mainly used during cooperation mode analysis.

\subsection{Experimental Settings}
There is only one important parameter, the cutoff distance $d_c$, which can be set by experience so that the number of nodes in each node's $d_c$ neighborhood is 1\%-2\% of the total number of network nodes~\cite{rodriguez2014clustering}. We calculate the distribution from 0.0 to 3.5 in the experiment and select the center node based on the observed value. The statistical results show that different  $d_c$ values have no obvious effect on the distinguishable distance of the cluster center but influence the $\delta$ of the cluster center. Therefore, when determining the cluster center, $\rho$ and $\delta$ are both standardized.  Tab.\ref{tab:impact} shows the impact impact of team recognition results in different $d_c$ values. We can see that $d_c$ within 1.5-2.5 has no obvious influence on the experimental results, and the robustness is good. In our experiment, $d_c$ is set as 1.6 for 2006-2009, 1.5 for 2008-2011, 1.5 for 2010-2013, 1.5 for 2012-1015, and 1.4 for 2014-2017.
\begin{table}[htbp]
  \centering
  
  \caption{The impact of team recognition result with $d_c$.}
  \resizebox{0.48\textwidth}{!}{
  \begin{tabular}{cccccccc}
    \toprule
    \textbf{$d_c$} & \textbf{0.5} & \textbf{1} & \textbf{1.5} & \textbf{2} & \textbf{2.5} & \textbf{3} & \textbf{3.5} \\
    \midrule
    2006-2009 & 14,566 & 14,867 & 14,988 & 14,988 & 14,988 & 14,593 & 14,354 \\
    2008-2011 & 21,946 & 22,003 & 22,279 & 22,280 & 22,280 & 22,014 & 22,003 \\
    2010-2013 & 27,017 & 27,127 & 27,628 & 27,628 & 29,628 & 27,319 & 27,278 \\
    2012-2015 & 27,143 & 27,274 & 28,207 & 28,207 & 28,205 & 27,920 & 27,674 \\
    2014-2017 & 28,179 & 28,380 & 29,530 & 29,530 & 29,530 & 28,739 & 28,240 \\
    \bottomrule
  \end{tabular}%
}
  \label{tab:impact}%
\end{table}%

\subsection{Evaluation Metrics}
To evaluate and analyze the effectiveness of MOTO for recognizing team results, we used five metrics: the number of recognized teams, team size, team communication cost, the number of triangle motifs, and separation degree. These metrics are introduced in detail below.

\textbf{Number of teams recognized and team size:} The number of recognized teams is one of the most basic metrics used. Team size is an important structural variable of a team, which can not be ignored. Appropriate team size is not only conducive to team communication, but also can improve team efficiency, which is the basic guarantee for completing research tasks of a scientific research team.

\textbf{Team communication cost:} It is an important indicator of whether or not the team cooperates effectively. The lower the communication cost, the more effective the team collaboration is. To measure this, we use Communication Cost Radius (CCR), i.e., the diameter of the induced subgraph of team members, which is the maximum length of the shortest path between any two nodes\cite{wang2016ustf,juarez2021comprehensive}. The calculation formula is shown in Eq.~\eqref{eq:ccr}.

\begin{equation}\label{eq:ccr}
CCR = \max _{{i,j} \in T}shortestPathLength(i,j)
\end{equation}

\textbf{Number of triangle motifs:} Triangle motifs are a connection mode that exist widely in social networks, which is also defined as triadic closure in social networks. 
 We use it as an indicator to evaluate the team structure. The more triangle motifs in the team, the closer the cooperation between team members.

\textbf{Separation degree:} It is used to measure the closeness of team members to the external and internal connections of the team. The greater the separation degree of the team, the closer the connection between team members and people outside the team; the smaller the separation degree, the closer the internal connection of the team. The measure can be calculated using the following equation


\begin{equation}
Separability(T)= \frac{Out_T}{All_T}
\end{equation}
where $Out_T$ is the number of connections between members of team $T$ and people outside the team. $All_T$ is the number of connections between members of team $T$ and the inside and outside of $T$.

\subsection{Baseline Approaches}
We use four methods as baseline approaches for comparison with our proposed algorithm: Team Recognition Algorithm based on CII (TRAC), Team Identification Based on iterative Centrality Ranking (TIBCR), Cluster Affiliation Model for Big Networks (BIGCLAM), and Discovering Community Cores (DCC). 

(1) \textbf{TRAC}~\cite{yu2017team} is a team identification algorithm based on the Collaboration Intensity Index (CII). It is a network edge weight filtering method. The first step is to set the edge weights in the collaboration network as CII. The second step is to screen network nodes according to Partnership Ability Index (PHI). The third step is to set the cooperation constraint coefficient $W$, delete the edge whose CII is lower than $W$, and delete the node without edges.

(2) \textbf{TIBCR}~\cite{Yu2018themethod} is a team leader and team identification algorithm based on iterative centrality ranking. The first step is to calculate and rank the intermediate centrality of each node in the academic collaboration network. Then, the 2-clique method is used to identify the core team members. Based on the team leader and core team members, the snowball method is used to identify the general team members.

(3) \textbf{BIGCLAM}~\cite{yang2013overlapping} is a model-based overlapping community detection method suitable for large networks. It can detect densely overlapping, hierarchically nested, and non-overlapping communities in massive networks. It first calculates the attribution vector of each node, and then uses a method based on matrix factorization to divide the community. The algorithm constructs a bipartite affiliation graph to simulate the structure of the community. Based on the new bipartite graph, it uses the graph adjacency matrix to maximize the affiliation matrix of the node.

(4) \textbf{DCC}~\cite{jones2016community} improves Speaker-listener Label Propagation Algorithm (SLPA). In SLPA, speaker-listener labels are allocated to different nodes according to the information transmission process. The labels are then spread among nodes according to the previous and current iteration information of the nodes. Finally, the labels are used to aggregate the nodes and form a community. DCC expands on this by setting the weight of the network to Intimacy.

TRAC is designed based on network edge weight filtering. This method is straightforward but neglects the team structure. Moreover, TRAC cannot recognize overlaps. TIBCR is an iterative method, which can respectively recognize team leader, core members, and other members. However, the iteration process is time-consuming. Therefore, it is not suitable for team recognition in large-scale networks. DCC improves the process of iteration to simplify the label update. Therefore, DCC is a more effective iterative method and meanwhile can recognize teams with overlaps. However, due to the randomness of label propagation, the recognition results of DCC are not stable. BIGCLAM considers community structure and membership strength to detect communities with overlap. This method utilizes coordinate ascending method to optimize non-negative matrix factorization. Therefore, it has obvious shortage in complexity, scalability, and linear model expression ability.

\subsection{Experimental Results}
In this section, we compare the result of our proposed algorithm with the baseline methods. Experiments were conducted using high-order familiarity (MOTO-H) and pairwise familiarity (MOTO-P), respectively.

\begin{figure}[tb]
  \centering
  \includegraphics[width=0.9\linewidth]{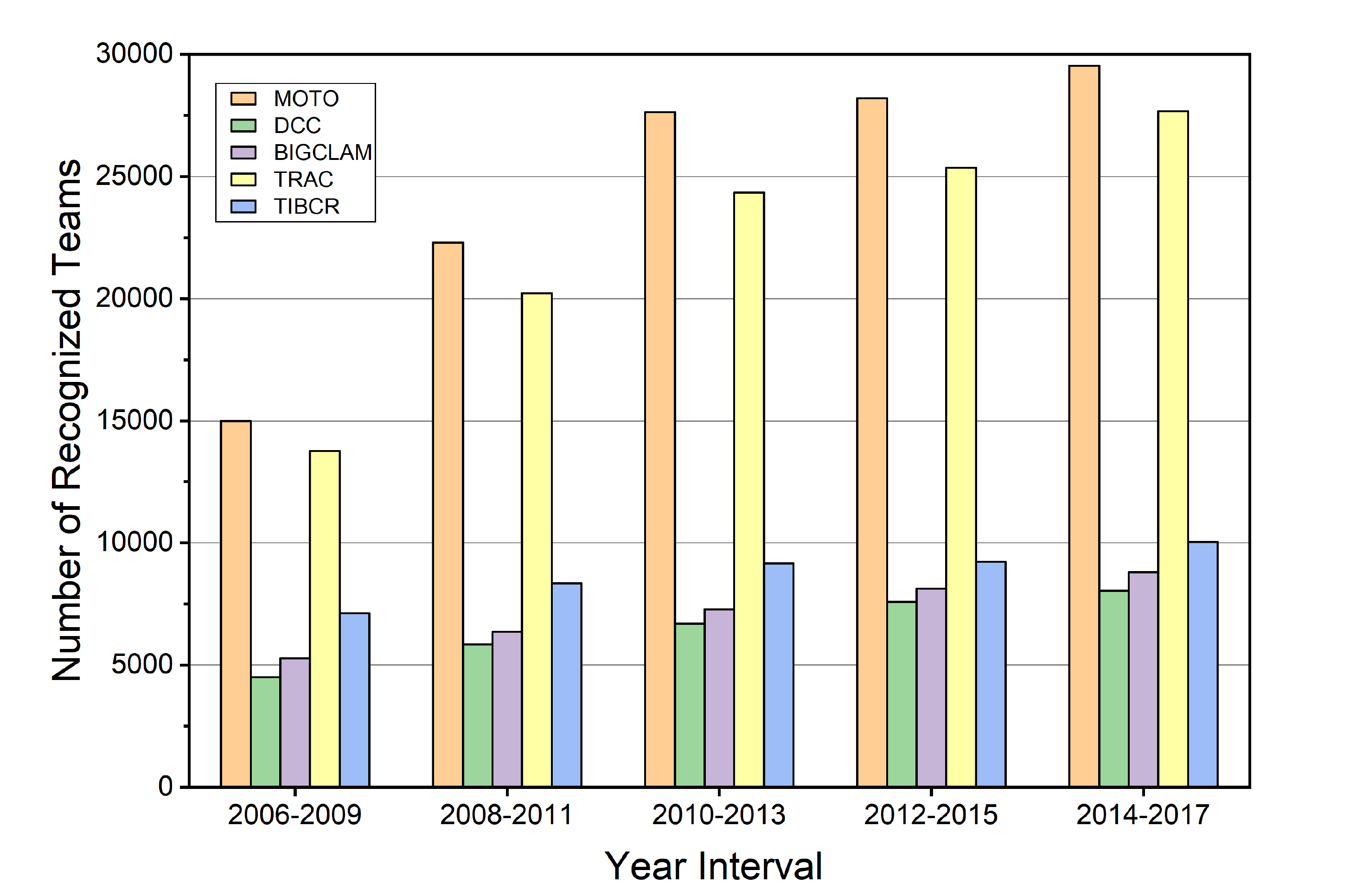}
  \caption{The number of recognized teams in time intervals.}\label{fig:recognized_teams}
\end{figure}

\textbf{Number of teams recognized and team size:} Fig.\ref{fig:recognized_teams} shows that the number of teams recognized by MOTO and comparison algorithm for different periods. The $x$-axis represents the period, and the $y$-axis represents the number of teams. This figure shows, that over time, the number of recognized teams increased. With respect to the overall network, the number of network nodes also increases over time, as shown in Tab.~\ref{tab:profile}. We determine that the size of the collaboration network has become larger in recent years, so the number of academic teams will also increase. This increase is consistent with our real world expectations. In all time periods, MOTO identifies the greatest number of academic teams when compared with baseline methods chosen for comparison.
\begin{figure}[tb]
  \centering
  \includegraphics[width=0.9\linewidth]{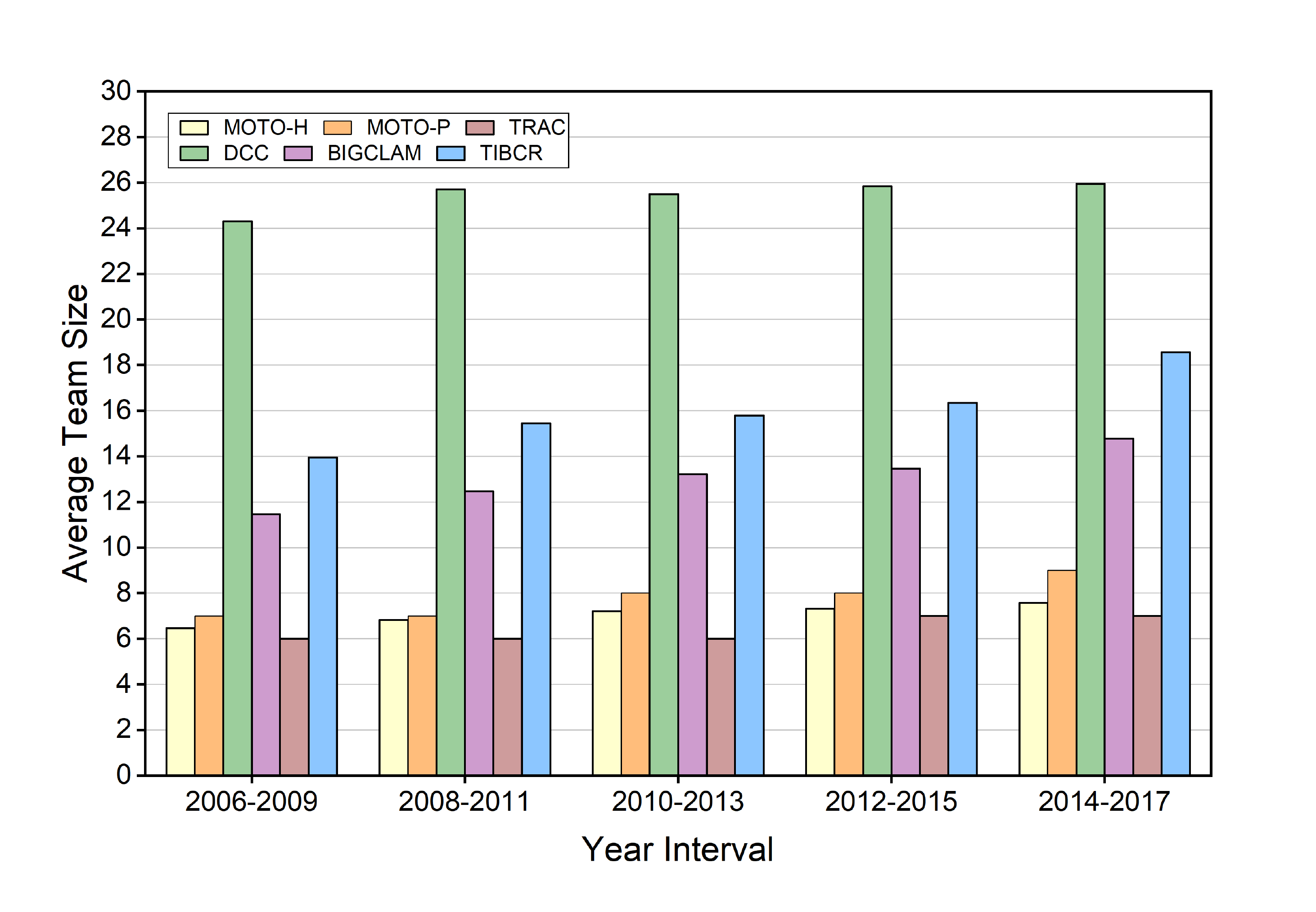}
  \caption{The average team size in time intervals.}\label{fig:teamsize}
\end{figure}

Fig.\ref{fig:teamsize} shows that the size of team recognized by MOTO-H and MOTO-P is the second and third smallest among all algorithms, respectively. MOTO-H and MOTO-P team sizes range between 7 to 10 people approximately. Therefore, MOTO recognizes more teams, but meanwhile the teams recognized by MOTO are with regular number of team members. That is to say, in the recognition process, MOTO does not split the teams. The team size recognized is slightly different with different familiarity. The high-order familiarity requires the team members to establish more triangle motifs representing cooperative relationships with other members, which is smaller than the team size obtained by pairwise familiarity. The number of teams recognized by the TRAC is also small but the difference to MOTO is less than about 14\%. The teams identified by TRAC are smaller, with approximately 6 or 7 members. The recognition results are closer to the average degree of nodes. However, the total number of teams is not the highest of all methods, because some nodes will become isolated nodes while deleting edges. DCC recognizes the least number of teams and the average size of the team is large, with 24-26 people, which does not match expectations based on real world observations. The team size recognized by TIBCR and DCC is between 11-19 members, which is a medium-sized team. The reason for the larger team size under TIBCR and DCC is that these two methods do not remove some members who do not cooperate closely.

In summary, the recognition result of MOTO is closer to our expectations of the team cooperation situation in reality. MOTO also loses minimal information when compared to other baseline methods. Choosing a higher level of familiarity will identify smaller teams. Over time, the size of the teams gradually increases.

\begin{figure}[tb]
  \centering
  \includegraphics[width=0.9\linewidth]{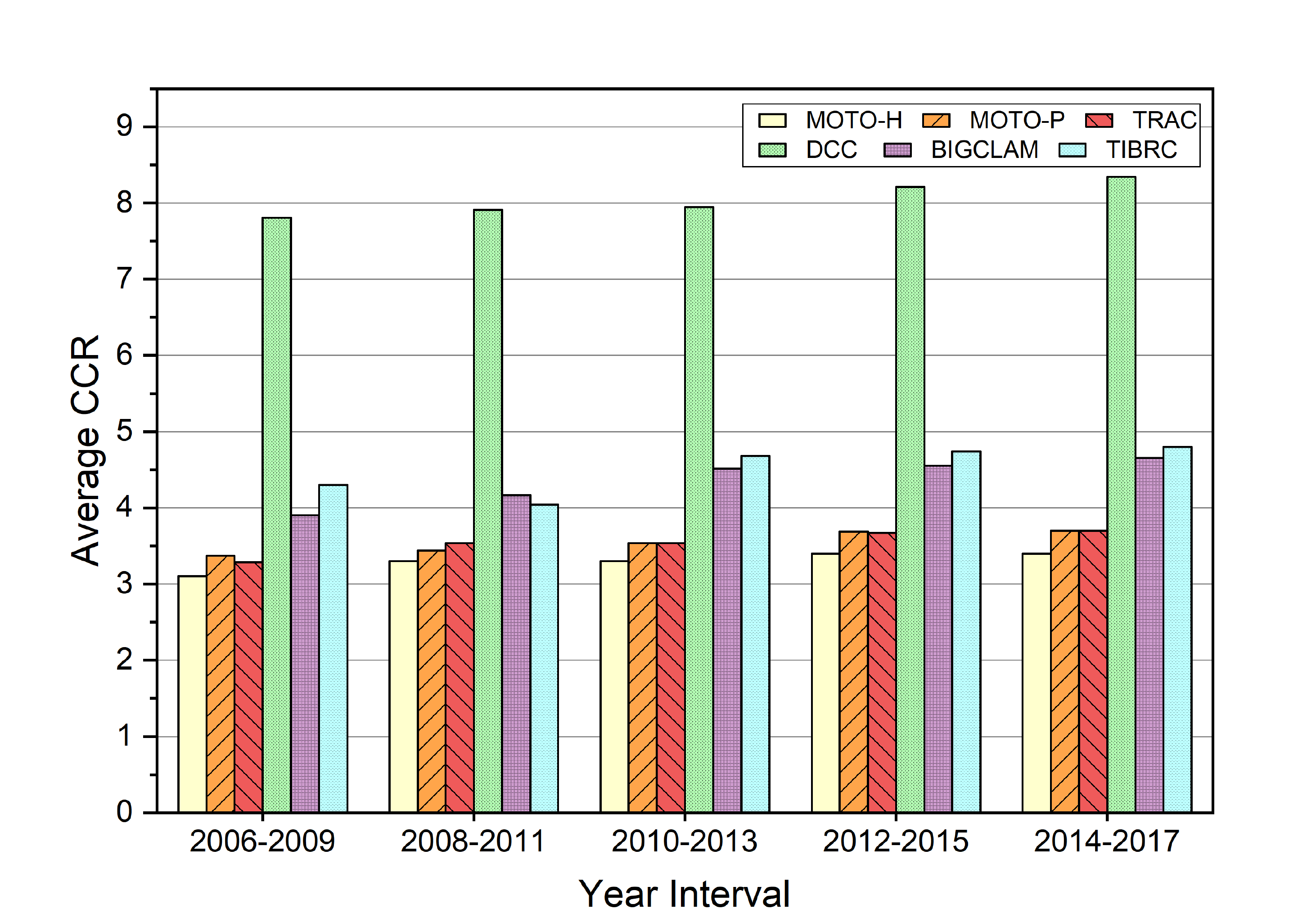}
  \caption{The average CCR of teams in time intervals.}\label{fig:ccr}
\end{figure}
\textbf{Team communication cost:} Fig.\ref{fig:ccr} shows the average CCR in each time interval. In each time interval, the results of MOTO-H and MOTO-P are the two lowest apart from the average CCR of MOTO-P in 2006-2009 which was less than 0.1 higher than TRAC. Overall therefore the performance of MOTO is similar to or better than TRAC. Similarly, the average team size recognized by MOTO-H and MOTO-P was larger than TRAC.  The communication costs of other algorithms are significantly higher than MOTO and TRAC. which is mainly because the size of the teams recognized by these three algorithms is significantly larger than MOTO and TRAC. It can be concluded that the teamwork recognized by MOTO-H and MOTO-P is more efficient than other methods. The difference between the CCR of the two MOTO algorithms is minimal (less than 0.3 across each time interval). We suggest that this is due to the team size identified by MOTO-H being slightly smaller than MOTO-P.  These results also indicate that the cost of high-order familiarity relationship cooperation is lower.

\begin{figure}[tb]
  \centering
  \includegraphics[width=0.9\linewidth]{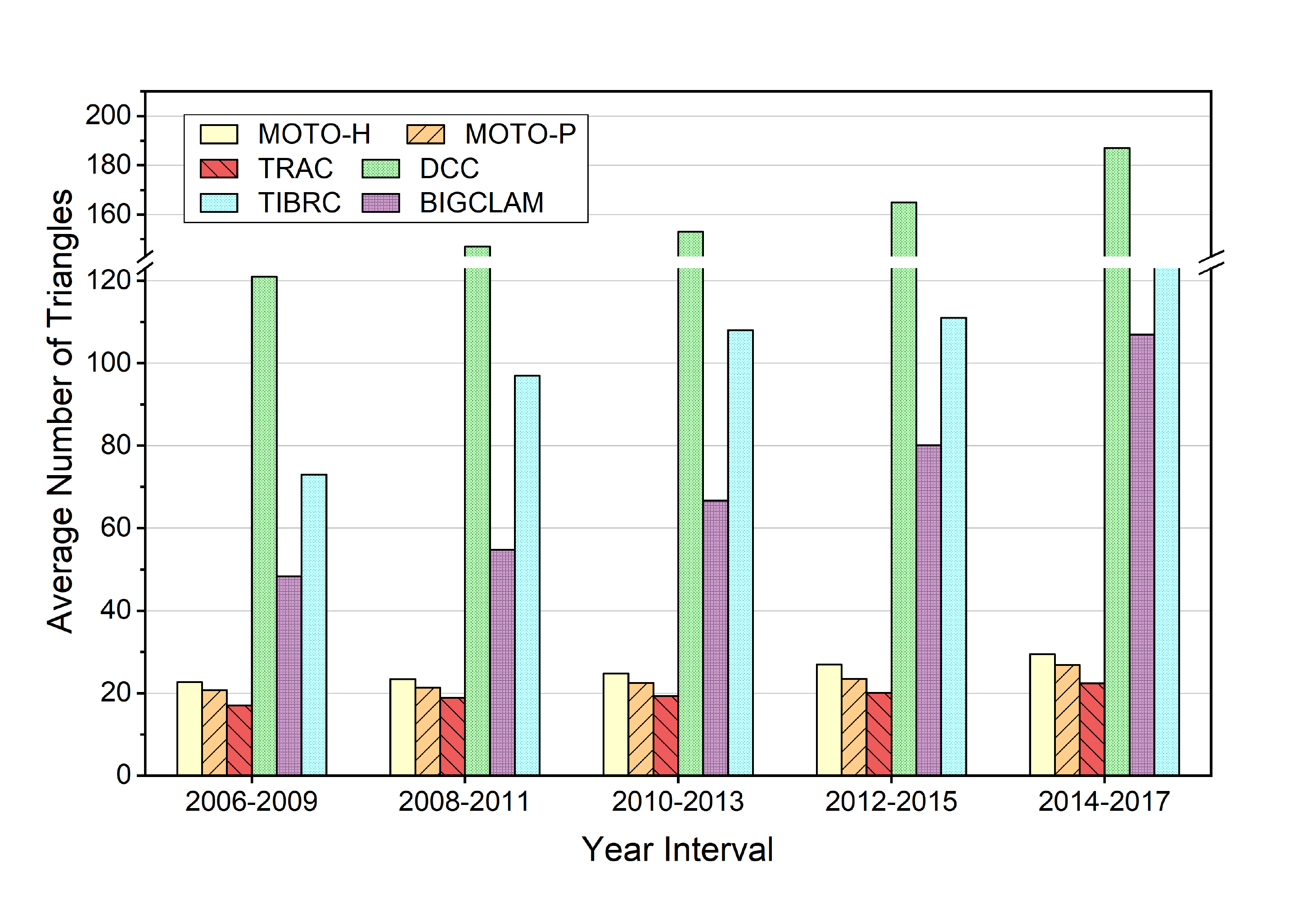}
  \caption{The average number of triangles in time intervals.}\label{fig:triangles}
\end{figure}

\textbf{Number of triangle motifs:} Fig.\ref{fig:triangles} shows the average number of triangle motifs recognized by different algorithms. The team sizes identified by DCC, TIBRC, and BIGCLAM are at least twice that of MOTO-H and MOTO-P, therefore more large triangle motifs will be present. Although the number of tree-order motifs in the team recognized by MOTO is far less than that of these three comparison algorithms, it does not mean that the team collaboration is not sufficiently represented. We comprehensively evaluated it by combining CCR and other indicators. When the team size differs by 1-3 people, the results of MOTO-H and MOTO-P are higher than TRAC by about 17\% and 26\%, respectively, which is not enough to exclude the influence of team size. Therefore, the recognition result of MOTO-H, MOTO-P, and TRAC is relatively reasonable in the angle of the triangle motifs number.

\begin{figure}[tb]
  \centering
  \includegraphics[width=0.9\linewidth]{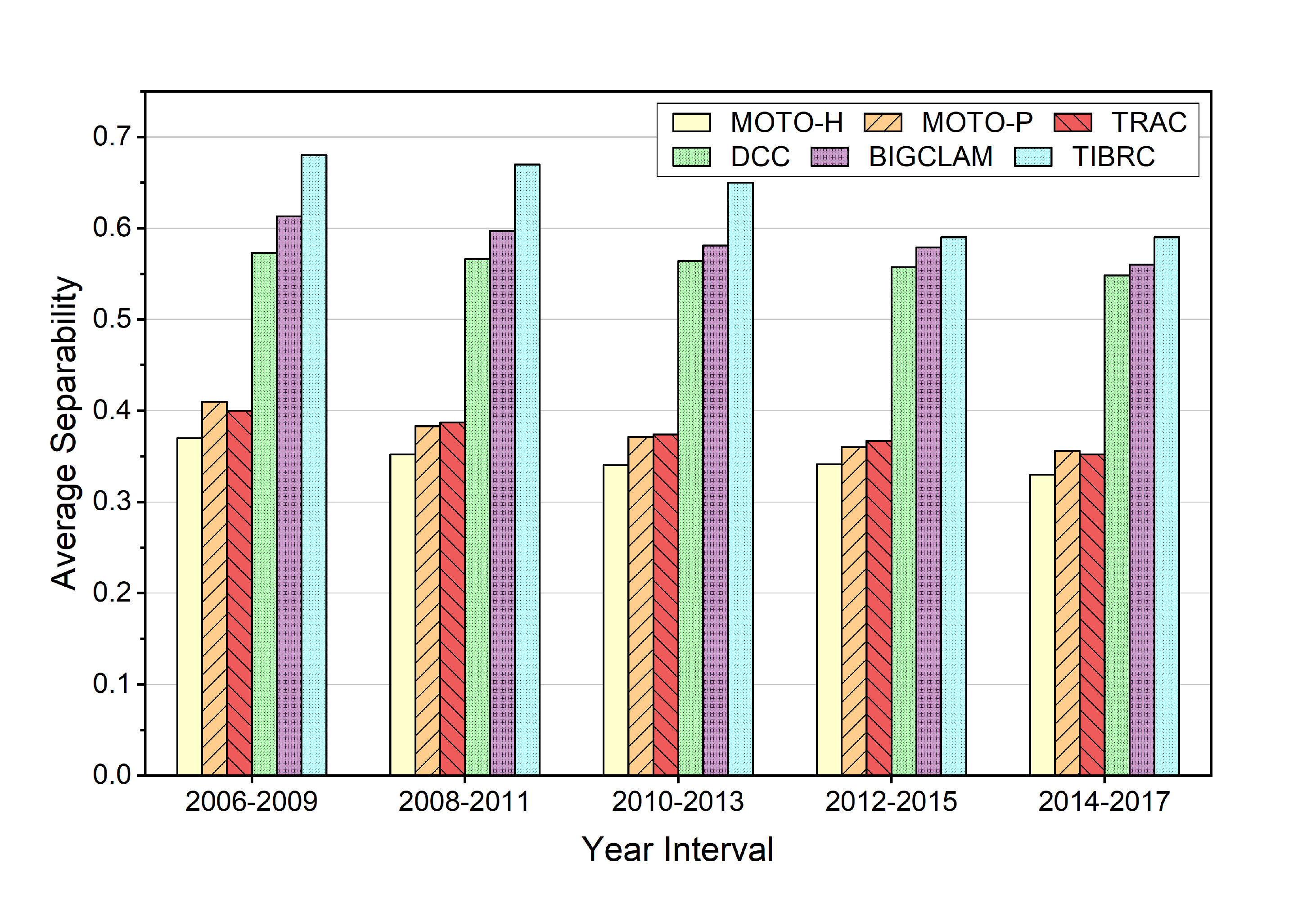}
  \caption{The average separability in time intervals.}\label{fig:separability}
\end{figure}
\textbf{Separation degree:} Fig.\ref{fig:separability} shows the average separation degree of each algorithm. The separation degree of DCC, BIGCLAM, and TIBRC are significantly higher than the others, and they are all greater than 0.5. The separation degree greater than 0.5 means that more than half of the team members' connections are connected to nodes outside the team, i.e., the connections inside the team are not tighter than those outside the team, so the recognized team structure does not match the real structure. Comparing MOTO-H, MOTO-P and TRAC, it can be found that the separation degree of MOTO-H is the lowest. The results of MOTO-P and TRAC are similar, but MOTO-P is generally lower.

Based on the above evaluations, MOTO recognizes the largest number of teams. Previous studies have illustrated that real team size varies from 3 to 8 \cite{yu2019academic} in the computer science discipline. The team sizes are closer to real team sizes, and there are fewer lost network structures. Comparing the results of high-order familiarity and pairwise familiarity, we find that the team that uses high-order familiarity has lower communication costs and a tighter structure. However, considering the higher computational complexity of high-order familiarity, we should select appropriate familiarity according to the actual situation for efficient academic team recognition.

\subsection{Analysis of Cooperation Model}
\begin{figure*}[ht]
  \centering
  \includegraphics[width=0.90\linewidth]{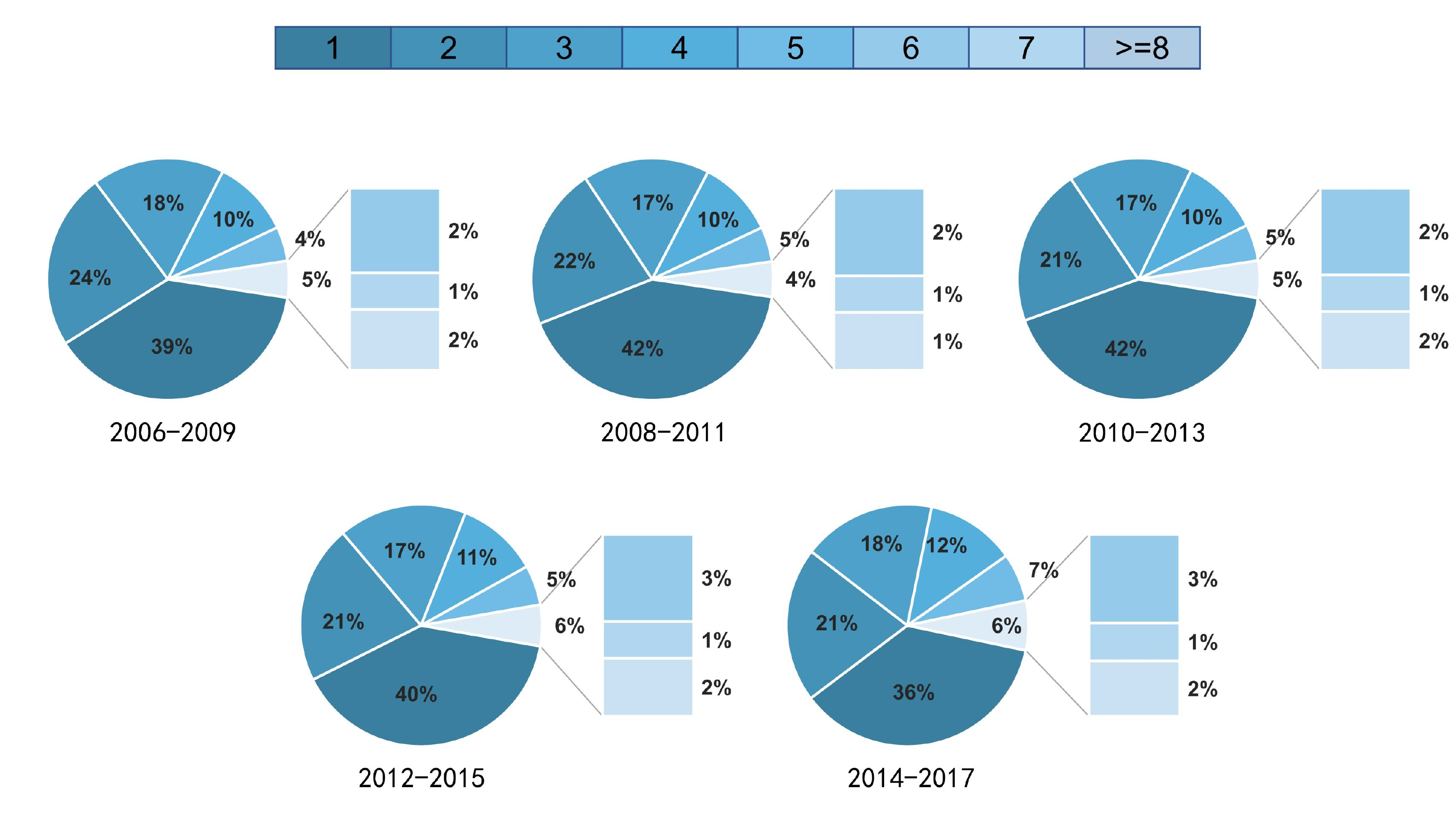}
  \caption{The number of authors who collaborate in one paper.}\label{fig:collaborate}
\end{figure*}
This section analyzes the cooperation mode and team performance of academic teams in combination with academic institutions. Academic teams can be divided into interagency teams and intra-institutional teams. 
According to the indicator evaluation in the previous section, we choose MOTO-H for analysis, because it has the best recognition result.

Firstly, we analyze the trend of scholars' cooperative behavior. Fig.\ref{fig:collaborate} shows the number of authors who collaborate on one paper. The number of the legend refers to the number of authors collaborated in one paper. In the legend, the darkest colored segment numbered with ``1" refers to the proportion of scholars who authored a paper alone. Likewise, the lighter the color is, the more collaborators in one particular paper are. The specific portion is correspondingly shown on the pie charts. For example, there are totally 39\% papers published with individual author during 2006-2009. According to these statistics, more than 60\% of the papers are completed cooperatively, and the proportion of co-authored papers is generally higher as time passes. The percentage of cooperation on papers is 3\% higher in 2014-2017 compared to 2008-2011. Simultaneously, the number of co-authors also increased over time, which suggests that the team size increases with time in the above recognition results.
\begin{figure}[tb]
  \centering
  \includegraphics[width=0.9\linewidth]{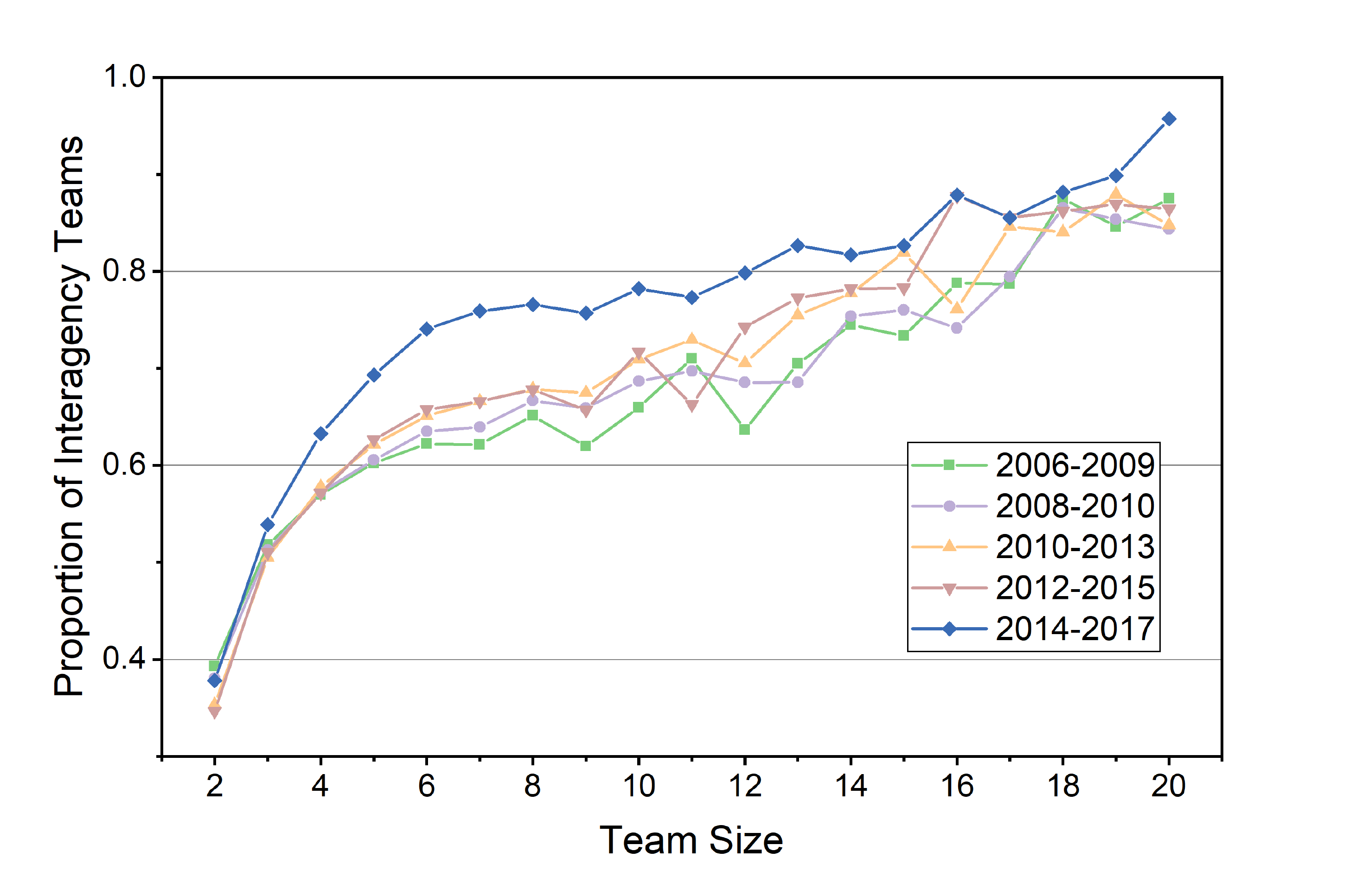}
  \caption{The proportion of interagency teams in academic teams.}\label{fig:result1}
\end{figure}
Secondly, we analyze the proportion of interagency teams. Fig.\ref{fig:result1} shows the proportion of interagency teams in academic teams with different team sizes in each period. When the team size is more than 20, the proportion of interagency teams is more than 83\%. However statistically, the number of teams with size greater than 20 is few and it is therefore difficult to use teams of this size for comparison purposes. Therefore, we select the team size of 2-20 for comparison. When the team size is 2, the proportion of interagency teams is about 30\% in all periods, which indicates that when the team has only two people working together, most of them are scholars from the same institution. When the team size is 3, the proportion is about 55\%. Teams of 4-8 people accounted for about 79\% of the recognition results, and the proportion of interagency teams exceeds 57\%. The larger the team size, the higher the proportion of interagency teams. This shows that interagency cooperation has become the main cooperation mode of teamwork. We also can see that when the team size is the same, the proportion of interagency teams has increased significantly over time. In particular, from 2014 to 2017, the proportion of interagency teams was significantly higher than in other periods by more than 10\%. When the team size reaches 16 or more, more than 80\% of the teams in all time periods involve interagency cooperation.

Finally, we analyze the impact of the increase of interagency academic teams on team performance. The team’s performance can be measured by the team’s average citations. The calculation formula is:

\begin{equation}
  {Citation}_{T}=\frac{\sum_{i \in T}  {citation}_{i}}{n_{T}}
\end{equation}
where  ${citation}_{i}$ is the sum of paper citations of $i$ in a certain period. $n_T$ is the team size.

\begin{figure}[tb]
  \centering
  \includegraphics[width=0.9\linewidth]{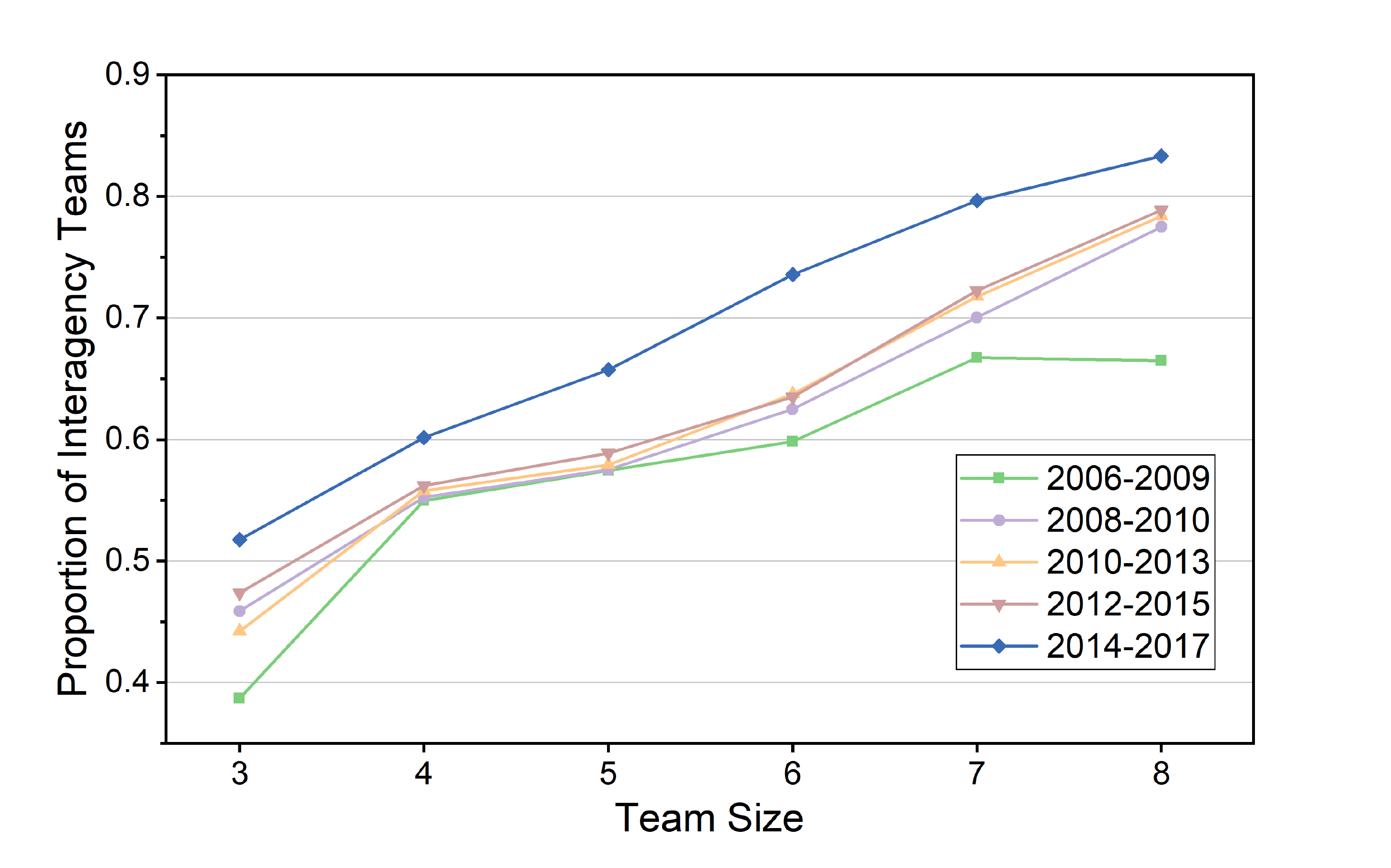}
  \caption{The proportion of interagency teams in academic teams with top-20\% cited.}\label{fig:result2}
\end{figure}

We sort the teams with the same team size in each period in descending order according to the average number of team references. In order to facilitate comparison, we select the top 20\% of teams with a team size of 3-8, as shown in Fig.\ref{fig:result2} since this selection provided a significant dataset for comparison purposes. When the team size is 3, the performance is relatively low in different time periods. When the team size is 4-6, the proportion of interagency teams and intra-agency teams is basically equal, i.e. their performance is comparable. When the team size is 7-8, the proportion of interagency teams is higher than that of intra-agency teams, i.e., interagency teams perform better than intra-agency teams.

\section{Conclusion}\label{sec:6}

Collaborative teams are assembled to fill the knowledge gap in academia better. There are a large amount of scientific research problems that demand solutions based on collaborative team work. Multi-variate factors including but not limited to familiarity, ability, team scale, and team composition together have an impact on the output of a team. How to optimize team structure, arrange resources, as well as enhance collaboration, are all fundamental issues that are needed to be solved. Therefore, in the beginning, collaborative teams should be firstly recognized to support continuous studies. In this work, we employ pairwise familiarity and higher-order familiarity to recognize collaborative teams in academia. Our proposed approach MOTO significantly outperforms baseline methods in a real-world, large-scale network. Teamwork patterns are also analyzed. Teams with members from different institutions widely exist in academia and generally achieve better performance. The number of teams also has an influence on team outputs. Our work provides a way to mine a large number of collaborative teams, which considers both collaboration behaviors and preferences. The proposed method MOTO can also be applied in other disciplines that feature abundant collaboration relationships. Considering the mobility in academia, the definition of familiarity will be optimized based on more data such as subjective consciousness or dynamic collaboration relations in our future work. We will also mine the recognized teams in-depth to identify and investigate new research patterns.

\bibliographystyle{ACM-Reference-Format}
\bibliography{reference}


\end{document}